\documentclass[12pt]{article}

\newcommand{\bbr}{I\!\! R}

\newcommand{\m}{\mathrm}
\newcommand{\be}{\begin{equation}}
\newcommand{\ee}{\end{equation}}
\newcommand{\ba}{\begin{eqnarray}}
\newcommand{\ea}{\end{eqnarray}}
\newcommand{\dif}{\mathrm{d}}

\usepackage{graphicx}
\usepackage{amssymb}
\usepackage{amsmath}
\usepackage[T1]{fontenc} 
\usepackage[ansinew]{inputenc} 
\usepackage[nosort]{cite}
\newcommand{\inbar}{\vrule height1.57ex width.4pt depth0pt}
\newcommand{\SW}{\relax{\hbox{$\ \inbar\kern-.285em{\rm S}$}}}

\begin{document}
\thispagestyle{empty}
\begin{center}

\null \vskip-1truecm \vskip2truecm

{\Large{\bf \textsf{Generalised Planar Black Holes and the Holography of Hydrodynamic Shear}}}

{\Large{\bf \textsf{}}}

{\large{\bf \textsf{}}}

{\large{\bf \textsf{}}}

\vskip1truecm

{\large \textsf{Brett McInnes $\,$and$\,$ Edward Teo
}}

\vskip0.1truecm

\textsf{\\ National
  University of Singapore}
  \vskip1.2truecm
\textsf{email: matmcinn@nus.edu.sg, eteo@nus.edu.sg}\\

\end{center}
\vskip1truecm \centerline{\textsf{ABSTRACT}} \baselineskip=15pt

\medskip
AdS black holes with planar event horizon topology play a central role in AdS/CFT holography, and particularly in its applications. Generalisations of the known planar black holes can be found by considering the Pleba\'nski--Demia\'nski metrics, a very general family of exactly specified solutions of the Einstein equations. These generalised planar black holes may be useful in applications. We give a concrete example of this in the context of the holographic description of the Quark-Gluon Plasma (QGP). We argue that our generalised planar black holes allow us to construct a model of the internal shearing motion generated when the QGP is produced in peripheral heavy-ion collisions. When embedded in string theory, the bulk physics is in fact unstable. We find however that this instability may develop too slowly to affect the evolution of the plasma, except possibly for high values of the quark chemical potential, such as will be studied in experimental scans of the quark matter phase diagram.

\newpage
\addtocounter{section}{1}
\section* {\large{\textsf{1. Holographic Description of the Internal Motion of the QGP}}}
Collisions of heavy ions \cite{kn:bedangadas} are believed to produce a state of matter known as the Quark-Gluon Plasma or QGP. One theoretical approach to understanding this state is based on \emph{holography}, in which the QGP is modelled by a field theory dual to a gravitational system, a thermal AdS black hole \cite{kn:solana,kn:pedraza,kn:youngman,kn:gubser}. Because the QGP exists in Minkowski space, one needs to use black holes with topologically planar event horizons; fortunately, these do exist in the asymptotically AdS case \cite{kn:lemmo}, though, as we shall see, these objects differ, in many important particulars, from their counterparts with topologically spherical sections.

The QGP, as produced in collisions, is far from being a static system: most obviously, it expands very rapidly, and much work is currently devoted to finding a dual description of the expanding QGP (see \cite{kn:gubser,kn:arefeva,kn:casalderrey,kn:craps1,kn:craps2} and references therein). However, it has recently come to be appreciated that, in non-central (``peripheral'') collisions, the QGP is also subjected to an extremely intense \emph{shearing} motion, because of the transverse non-uniformity of the colliding nuclei \cite{kn:liang,kn:bec,kn:huang}. The large vorticity thus generated \cite{kn:bemo,kn:leigh1,kn:leigh2,kn:leigh3,kn:sorin} may prove to have observable consequences, through quark polarization, and it may be important for forthcoming attempts to simulate peripheral collisions using holography \cite{kn:romat}.

Clearly it is important to determine what the holographic approach has to teach us about this shearing effect. Since the shearing imparts a very large angular momentum to the plasma, the natural suggestion \cite{kn:schalm} is to consider AdS-Kerr \cite{kn:carter,kn:hawrot} black holes in the bulk. However, while the AdS-Kerr black hole (with a topologically spherical event horizon) does induce rotation at infinity, the angular velocity is constant (that is, independent of direction) there, and hence there is no shear. This spacetime is indeed of interest for possible holographic descriptions of systems which are genuinely \emph{rotating}
(see \cite{kn:sonner} for an explicit example, and \cite{kn:rotation} for another potential application), but this is not what we seek here.

In any case, the AdS-Kerr metric does not have a conformal boundary which is globally conformal to Minkowski spacetime, so one should really begin rather with non-static black holes which do have that property ---$\,$ that is, with planar black holes possessing angular momentum. Such black holes were discovered by Klemm, Moretti and Vanzo \cite{kn:klemm}, and one finds that, in addition to inducing a more reasonable geometry at infinity, these ``KMV$_0$ spacetimes'' bring with them a major advantage: the angular velocity at infinity depends on one of the spatial coordinates, \emph{thus inducing an effective shearing motion on the boundary}. The KMV$_0$ spacetimes, and their electrically charged (``QKMV$_0$'') versions, are therefore prime candidates for building a holographic description of the shearing QGP. Such a description may be very important, because it often happens that certain effects are apparent on one side of a holographic duality, but obscure on the other. Work in this direction was described in \cite{kn:75,kn:76}.

The shearing motion of a fluid is described by a \emph{velocity profile}, an expression of the (dimensionless) velocity $v(x)$ as a function of transverse distance from some axis. (In the case of plasma generated by heavy-ion collisions, it is customary to choose the axis to be that of the collision, that is, the axis along which the velocity vanishes; it is conventional to take it to be the $z$-axis.) This function is of basic importance, since it describes the internal dynamics of the plasma. Unfortunately, in the extreme conditions in the aftermath of a heavy-ion collision, it is difficult to predict the \emph{precise} shape $v(x)$ actually takes; however, typical shapes arising in shear flows are known (see \cite{kn:drazin}, page 133). Such a shape, chosen because it is consistent with causality, is shown in Figure 1.

\begin{figure}[!h]
\centering
\includegraphics[width=0.55\textwidth]{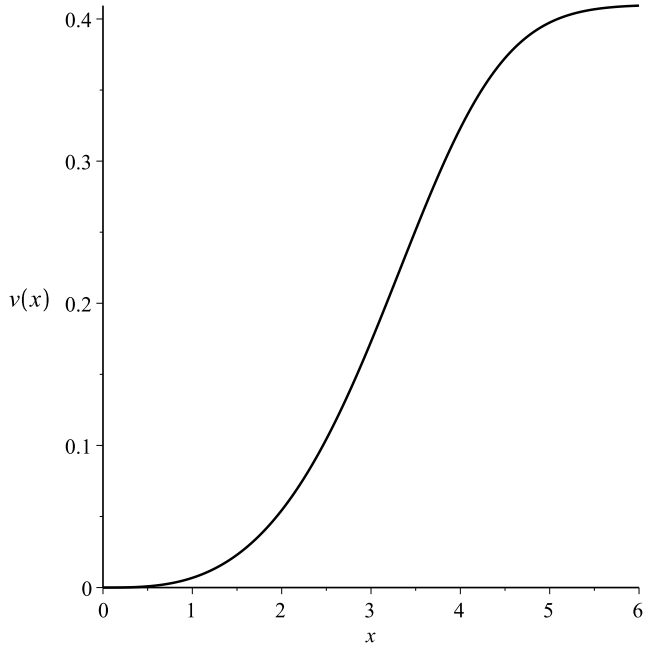}
\caption{Velocity profile of an unstable flow.}
\end{figure}

Crucially, some aspects of fluid behaviour actually depend only on the \emph{general shape} of the profile, not on the details; these aspects, therefore, can be discussed in the QGP case, even in the absence of a precisely specified $v(x)$ function. Most important among these is the question as to whether the shearing motion described by $v(x)$ is stable, in the hydrodynamic sense. Frequently it is not, the most well-known example being the \emph{Kelvin--Helmholtz instability}, which is important in numerous fluid mechanics problems where shear causes a laminar flow to become turbulent. Remarkably, all one needs to detect this instability, in the case of a monotone profile, is a knowledge of the \emph{sign} of the second derivative of $v(x)$. To be precise, a necessary (and usually sufficient) condition for Kelvin--Helmholtz instability to be present is that there should exist a value of $x$ such that the following is satisfied:
\begin{equation}\label{FJORTOFT}
{\dif^2 v\over \dif x^2}(x)\,\big[v(x)\,-\,v(x_i)\big]\;<\;0,
\end{equation}
where $x = x_i$ is the location of a point where the second derivative vanishes. (This is the classical ``Fj{\o}rtoft theorem'': see \cite{kn:drazin}, page 132.)

\begin{figure}[!h]
\centering
\includegraphics[width=0.55\textwidth]{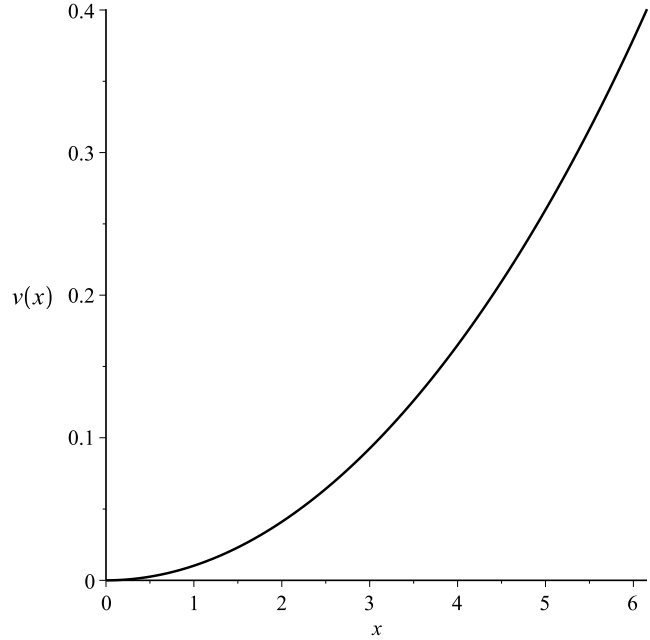}
\caption{QKMV$_0$ velocity profile; classically stable.}
\end{figure}
\begin{figure}[!h]
\centering
\includegraphics[width=0.55\textwidth]{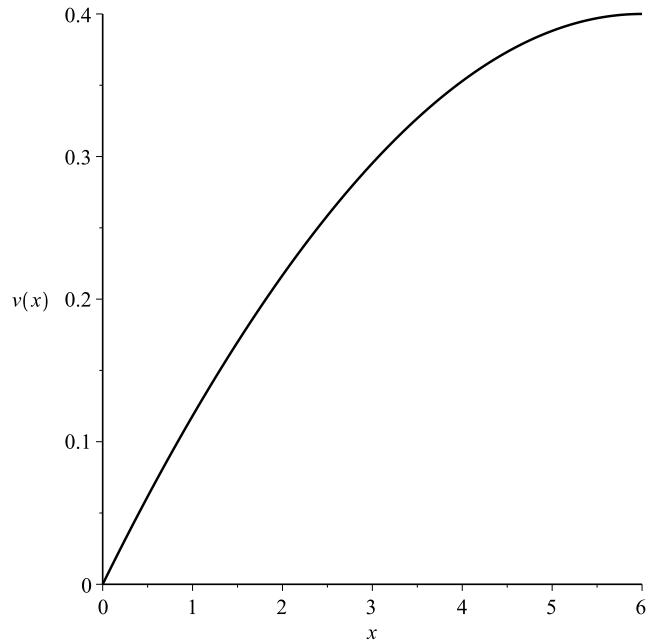}
\caption{$\ell$QKMV$_0$ velocity profile; classically stable.}
\end{figure}

Thus, for example, one should expect a fluid velocity profile like the one in Figure 1 to be unstable\footnote{Strictly speaking, the Fj{\o}rtoft theorem only gives a necessary condition for instability; however, it is thought on physical grounds (see again \cite{kn:drazin}, page 132) that the condition is also sufficient in the case of monotone profiles, which are the only kind relevant here.}, but a profile like the one in Figure 2 should be entirely stable (whether or not $x = 0$ is a point of inflection). Similarly, the profile in Figure 3 will be stable if there are no points of inflection in the graph (including at $x = 0$). Notice that the graph in Figure 1 can be constructed by taking the lower half of the graph in Figure 2 and combining it with the upper half of the graph in Figure 3 (the ranges of $x$ and $v(x)$ being the same in all three figures). Thus we see that the instability arises only in the middle of the flow, around the point of inflection in $v(x)$, never in regions with profiles having no such point.

We can summarise by saying that a detailed knowledge of the precise shape of $v(x)$ is not required to make these statements about the locations in which shear-induced hydrodynamic instability should be expected to arise: it will suffice if we can understand the behaviour of fluids having velocity profiles with a given general shape (concave ``up'' or ``down''). The physical intuition \cite{kn:lin} here is simple (but very classical): if a fluid cell with large vorticity (such as those near a point of inflection in $v(x)$) is moved slightly towards a region of lower vorticity, it will not tend to return to its original position.

It is clearly very important to understand whether shear-induced instabilities can affect the flow of the QGP arising from peripheral collisions. In a recently proposed theoretical model of the shearing QGP (see \cite{kn:KelvinHelm,kn:viscous} and references therein), it is suggested that the velocity profile does take a form satisfying the conditions of Fj{\o}rtoft's theorem, and simulations of the consequent Kelvin--Helmholtz instability are generated. This may well lead to observable effects, and so one obtains, in principle, a way of testing the model.

This discussion is based, however, on assuming that the QGP is fully governed by classical hydrodynamics. In particular, it is assumed in \cite{kn:KelvinHelm,kn:viscous} that the instability is localised, as above, around points of inflection in the velocity profile. But it is not entirely clear that the QGP is governed by purely classical effects; in view of the many forms of instability well known to be associated with other plasmas, one should be prepared to encounter deviations from classical expectations in this case also. This is precisely the kind of question on which the holographic approach might be expected to throw some light, and, in fact, the theory of holographic corrections to classical hydrodynamics is currently a subject of great interest \cite{kn:oog}.

In this work, we will use the QKMV$_0$ geometries, and generalizations of them, to construct the AdS/CFT duals of boundary configurations corresponding to classically stable velocity profiles of the kinds portrayed in Figures 2 and 3. The objective is to study whether the classical stability continues to hold when the bulk physics is considered.

The QKMV$_0$ metrics actually induce velocity profiles at infinity of the form given by Figure 2, but never of the kind given in Figure 3. They are therefore unable to give a complete description of a realistic profile (since such a profile would certainly, by causality, correspond to a bounded function, so at least part of the full graph must be of the form given in Figure 3). However, Klemm et al.~\cite{kn:klemm} obtained their metric as a special case of the Pleba\'nski--Demia\'nski family of metrics \cite{kn:plebdem,kn:grifpod}, the most general metrics of Petrov Type D \cite{kn:exact}, and therefore the most general spacetimes of black hole type that one is able to examine explicitly. The idea suggests itself that a metric inducing a profile like the one in Figure 3 may also be found as a more general member of that family. The suggestion that some of the less familiar Pleba\'nski--Demia\'nski metrics may be physically relevant has indeed been explored recently \cite{kn:sparks,kn:nozawa}. The results are encouraging, in that they indicate that these metrics have good properties from a supersymmetric point of view, and thus might ultimately be embedded in string theory.

In this work we exhibit a generalised planar black hole geometry of this kind. This geometry is characterised by its ADM mass, electric charge, angular momentum parameter, and by a new parameter conventionally denoted by $\ell$. The physical meaning of this parameter is, in general, obscure\footnote{For topologically spherical black holes, $\ell$ has the traditional interpretation of NUT charge, which would lead to closed timelike curves (CTCs) in the spacetime. Roughly speaking, this arises when the axis of symmetry has disconnected parts that are separated by the black hole. Without loss of generality, an angular coordinate can be identified with the standard periodicity on one part of the axis; however, NUT charge will present an obstacle to an analogous identification on the other part of the axis. A consistent identification will lead to a periodicity in the time direction as well, and hence to CTCs. For the planar black holes considered here, there is {\it no\/} such axis of symmetry, and hence none of the associated problems of NUT charge.}; but we argue that, in the application to heavy-ion holography, it has an unambiguous interpretation as a length scale determined by the physical parameters of the collision. The velocity profile at infinity is precisely as given in Figure 3 ---$\,$ that is, it could represent at least part of the velocity profile of the shearing QGP, the part far from the axis (in Figure 1). We show that the bulk geometry can be precisely specified in terms of observational data.

We find that a plasma with either of the profiles in Figures 2 and 3 is dual to an \emph{unstable} system in the bulk: this is due to a specifically ``stringy'' bulk effect, the instability associated with the production of branes wrapping around transverse sections \cite{kn:seiberg}. The corresponding instability in the fluid is not localised near a point of inflection in the velocity profile. It corresponds to a rather dramatic transition to turbulence \emph{throughout} the fluid.

However, one must always bear in mind that the QGP formed in any heavy-ion collision is inevitably a very short-lived state. We propose a holographic method of estimating the time required for our non-classical instability to develop, and find that, under the conditions of experiments performed hitherto, this time may well be too long to affect the plasma before it hadronises. If that is so, the idea that hydrodynamic instability arising from shear is localised in the manner proposed in \cite{kn:KelvinHelm} is valid for these experiments. The instability revealed by holography does however develop much more rapidly when the quark chemical potential is large; this is interesting, because plasmas with large chemical potentials are the subject of much experimental interest, in the ``beam energy scan'' programmes projected for such facilities as RHIC, SHINE, FAIR, and NICA \cite{kn:ilya,kn:dong,kn:shine,kn:fair,kn:nica}. It is possible that, under these conditions, the plasma will behave in a manner quite different from predictions based on a simple Kelvin--Helmholtz effect.

\addtocounter{section}{1}
\section* {\large{\textsf{2. The Boundary Geometry}}}
A simple argument, presented very clearly in \cite{kn:bec}, shows that, as a result of the non-uniformity of the distribution of nucleons in the transverse direction, the QGP produced in a peripheral heavy-ion collision acquires a very large angular momentum density associated with a non-trivial transverse velocity profile. Planar black holes with non-zero angular momentum, first described in \cite{kn:klemm}, do not actually ``rotate'': they are characterised by a \emph{shearing} motion in each transverse section strictly outside the event horizon. These planar black holes\footnote{Although these objects are often called ``black branes'', there is a risk of confusion with the actual physical branes that appear later in this work, and also with related higher-dimensional objects; therefore, to avoid misunderstandings, we shall refer to them exclusively as planar black holes.}, are therefore the basis for any holographic investigation of the shearing QGP.

More generally, a planar black hole will be of interest for this purpose if it induces at infinity a ``peripheral collision geometry'' with a metric of the form
\begin{equation}\label{A}
g_{\rm PC} \;=\; - \, \m{d}t^2 \;-\; 2\omega_{\infty}(x)\, L \, \m{d}t\m{d}z \;+\; \m{d}x^2 \;+\; \m{d}z^2.
\end{equation}
Here $x$ and $z$ are Cartesian coordinates describing a typical section through the plasma, with $z$ being the axis of the collision which produced that plasma; the function $\omega_{\infty}(x)$ is the asymptotic value of the angular velocity of the black hole (so we want this function to vanish along the $z$ axis, that is, for $x$ = 0). It depends on $x$, so that the velocity profile is not trivial; it also depends on various parameters of the underlying black hole, such as its angular momentum, and on $L$, the asymptotic AdS curvature scale.

Free particles in the spacetime described by the metric in (\ref{A}), with $x$ = constant and zero momentum in the $z$ direction, nevertheless move in the $z$-direction at a dimensionless speed given by
\begin{equation}\label{B}
v(x) \; \equiv \; \m{d}z/\m{d}t = \omega_{\infty}(x)L,
\end{equation}
so the function $v(x)$ describes the velocity profile at infinity. This motion is just the usual frame-dragging associated with \emph{any} black hole with angular momentum: but the essential difference between asymptotically flat and asymptotically AdS black holes is that, in the latter case, \emph{frame-dragging persists even at infinity}. This frame-dragging at infinity opens up the possibility of using the AdS/CFT duality to study rotating matter in the topologically spherical case \cite{kn:sonner}, and shearing matter in the topologically planar case.

Let us see how this works in the special case of the asymptotically AdS black hole geometry given by Klemm et al.~\cite{kn:klemm} (actually, we will use the electrically charged version, which we call the ``QKMV$_0$'' metric). That metric takes the form
\begin{equation}\label{C}
g({\rm QKMV}_0) = - {\Delta_r\Delta_{\psi}\rho^2\over \Sigma^2}\,\m{d}t^2\;+\;{\rho^2 \over \Delta_r}\m{d}r^2\;+\;{\rho^2 \over \Delta_{\psi}}\m{d}\psi^2 \;+\;{\Sigma^2 \over \rho^2}\big[\omega\,\m{d}t \; - \;\m{d}\zeta\big]^2,
\end{equation}
where the asymptotic AdS curvature is $-1/L^2$, where $\psi$ and $\zeta$ are dimensionless planar coordinates, and where
\begin{eqnarray}\label{D}
\rho^2& = & r^2\;+\;a^2\psi^2 \nonumber\\
\Delta_r & = & {r^4\over L^2} - 8\pi M^* r + a^2 + 4\pi Q^{*2}\nonumber\\
\Delta_{\psi}& = & 1 +{a^2 \psi^4\over L^2}\nonumber\\
\Sigma^2 & = & r^4\Delta_{\psi} - a^2\psi^4\Delta_r\nonumber\\
\omega & = & {\Delta_r\psi^2\,+\,r^2\Delta_{\psi}\over \Sigma^2}\,a.
\end{eqnarray}
Here $M^*$ and $Q^*$ are the mass and charge parameters, defined in terms of densities at the horizon (so that, for example, the charge density at the horizon is given by $Q^*/r^2_{\m{H}}$, where $r_{\m{H}}$, the value of $r$ at the horizon, is the largest root of $\Delta_r$; see \cite{kn:75} for this procedure, which is necessitated by the non-compact topology of the $r = constant$ sections here.) Similarly, the parameter $a$ is the ratio of the angular momentum and energy densities at the horizon.

Taking the $r \rightarrow \infty$ limit, and defining Cartesian coordinates by d$x$ = $L$d$\psi /\sqrt{1 + a^2 \psi^4/L^2}$ and d$z$ = $L$d$\zeta$, one finds \cite{kn:76} that the conformal boundary metric is given precisely by equation (\ref{A}) above, with asymptotic angular velocity function $\omega_{\infty}(x)$ defining a velocity profile given by
\begin{equation}\label{E}
v(x)\;=\;\omega_{\infty}(x)L\;=\;{1\over \gamma^2}\,\m{sinl}^2\Big(\sqrt{{a\over L}}\,\gamma\,{x\over L}\Big),
\end{equation}
with $\gamma$ = $(1 + i)/\sqrt{2}$, and where sinl($x) \equiv$ sn($x, i)$ is the ``lemniscatic sine'', one of the Jacobi elliptic functions \cite{kn:abramo}, sn($x, k$), with imaginary elliptic modulus. (This function is real, periodic, but not bounded here, since it is being evaluated off the real axis.)

Choosing typical parameter values, one obtains the QKMV$_0$ velocity profile, pictured in Figure 2. All of the velocity profiles obtained from QKMV$_0$ metrics are of this shape. Notice that there is no point of inflection here, even at the origin; this can easily be shown directly from the properties of the Jacobi elliptic function (or by using equation (\ref{I}), below). Thus a fluid with this profile is classically stable.

We are therefore in a position to give a holographic account of fluids with profiles like those in Figure 2. Our task is now to find some new bulk geometry such that, in a similar manner, a profile of the form shown in Figure 3 can be obtained. We shall do that in the next section, but first we study the (surprisingly strong) constraints imposed by the physics of the boundary field theory.

We begin by noting that the QKMV$_0$ boundary metric is not flat, but (as we will see) it \emph{is} conformally flat, like the boundary of the topologically spherical AdS-Kerr-Newman spacetime, which is used to study rotating superconductors \cite{kn:sonner}. Unlike the latter, it actually is \emph{globally} conformally flat\footnote{By this we mean that the conformal factor which reduces the metric to a flat metric on $\bbr^n$ does not vanish anywhere.
To show this, we complexify and so transfer the problem to the Euclidean domain. We then refer to the proof of Kuiper's theorem \cite{kn:kuiper} in conformal geometry, which states that a locally conformally flat, compact, simply connected Riemannian manifold must be globally conformal to the round sphere. In the course of the proof, before compactness is used, one shows that, given a locally conformally flat, simply connected $n$-dimensional manifold $M$, with the topology of $\bbr^n$, there exists a smooth globally conformal map from $M$ to an open proper subset of the sphere, $S^n$. By stereographic projection, which is itself a global conformal map, one can therefore show that $M$ is globally conformal to a piece of $\bbr^n$ with its standard flat metric.}, which means that the conformal fields on the boundary effectively propagate on Minkowski space. This is what we want, since the actual QGP produced in colliders exists, of course, in flat spacetime.

As we try to generalise the QKMV$_0$ metric, then, we should do so in a way that preserves the conformal flatness of the boundary. Let us investigate the conditions under which a general metric of the form given in equation (\ref{A}) can be conformally flat. The condition for this \cite{kn:cotton} is the vanishing of the \emph{Cotton tensor},
\begin{equation}\label{G}
C(g)= \eta^{mn(i}\nabla_mR_n^{j)}\,\partial_i\otimes \partial_j,
\end{equation}
where $\eta^{ijk}$ is the alternating tensor and $R_{ij}$ is the Ricci tensor. It is clear that requiring this object to vanish for the metric in equation (\ref{A}) will impose a third-order differential equation on $\omega_{\infty}$; the equation has the form
\begin{equation}\label{H}
\omega_{\infty}'''(\omega_{\infty}^2L^2 + 1)^2\;-\;4L^2(\omega_{\infty}^2L^2 + 1)\omega_{\infty}\omega_{\infty}'\omega_{\infty}''\;+\;L^2(3\omega_{\infty}^2L^2 - 1)(\omega_{\infty}')^3 \;=\;0,
\end{equation}
where the dash denotes a derivative with respect to $x$. Since $x$ does not appear explicitly, one constant of integration will be trivial (constant translations of $x$); the other two will be the physically significant ones.

Surprisingly, this complicated equation can be reduced to a first-order equation: see \cite{kn:76} for the details. One finds that all solutions of equation (\ref{H}) are either constant or solutions of the equation
\begin{equation}\label{I}
{1\over 2}(\omega_{\infty}')^2L^4\;=\;(A\,+\,B\omega_{\infty} L)(1\,+\,\omega_{\infty}^2L^2),
\end{equation}
where $A$ and $B$ are dimensionless constants. Because $\omega_{\infty}$ should vanish at $x = 0$, we see that $A = (1/2)\omega_{\infty}'(0)^2L^4$, so $A$ is either zero or positive; similarly $B$ must be positive if $A = 0$; otherwise $A$ and $B$ are arbitrary: they are the two constants mentioned above.

The function sinl$^2\Big(\sqrt{B\over 2}\,\gamma\,{x\over L}\Big)/(\gamma^2L$) satisfies equation (\ref{I}) with $A = 0$, and so, if we set $B = 2a/L$ (which means that the parameter $a$ has to be positive in this case), we get exactly the expression in equation (\ref{E}). This confirms that the QKMV$_0$ boundary is conformally flat (and it gives us the physical interpretation of the constant $B$, in terms of the angular momentum parameter of the underlying planar black hole).

On the other hand, setting $B$ = 0, we obtain an equation which can be solved in terms of an elementary function:
\begin{equation}\label{J}
\omega_{\infty}\;=\;{1\over L}\,\m{sinh}\Big(\sqrt{2A}\,{x\over L}\Big).
\end{equation}
The graph of this function is similar to the one portrayed in Figure 2, so this geometry, too, describes the part of the graph in Figure 1 with small $x$. This case does have one unique feature: the graph, like that of a realistic profile, does have a point of inflection at $x = 0$. However, the condition in Fj{\o}rtoft's theorem is not satisfied even here, so, once again, a fluid with this profile is classically stable.
It is easy to see that similar results are obtained if both $A$ and $B$ are positive: all of the graphs will resemble Figure 2. All of these geometries ($A = 0$, $B >0$; $A > 0$, $B = 0$; and $A > 0, B > 0$) are candidates to describe that kind of profile.

One possibility remains: that $A$ should be positive, and $B$ negative. Clearly the slope of the graph in this case gradually decreases, until a maximum is reached. If we cut off the solution at that point, then we can expect to obtain a graph like the one in Figure 3. This case therefore plays an essential role: apart from any other consideration, the velocity is bounded in this case (only), so causality is respected. This is the case on which we focus for the remainder of this section.

It will be convenient to adopt a new notation at this point. Let us use the velocity directly, instead of the asymptotic angular velocity, as in equation (\ref{B}). Next, replace $B$ by $-2|a|/L$, as above in our discussion of the QKMV$_0$ case, except that now the parameter $a$ must be negative. Finally, for reasons which will become clear, we wish to replace $A$ by 2$\ell^2/L^2$, where $\ell$ is a new parameter, with units of length, which remains to be interpreted\footnote{With this notation, equation (\ref{J}) becomes $\omega_{\infty}\;=\;{1\over L}\,\m{sinh}(2\,\ell x/L^2);$ clearly we want $\ell$ to be positive. See the next section for the dual geometry in this case.}. Then we have, from equation (\ref{I}),
\begin{equation}\label{K}
\bigg({\m{d}v\over \m{d}x}\bigg)^2\;=\;{4|a|\over L^3}\big(V - v\big)\big(1\,+\,v^2\big),
\end{equation}
where $V$ is a constant defined by
\begin{equation}\label{L}
\ell^2\;=\;V|a|L.
\end{equation}

It is clear from equation (\ref{K}) that $V$ has a simple physical interpretation: it is the maximal velocity of the plasma, some large fraction of the speed of light in the case of a heavy-ion collision. Inspection of (\ref{K}) also reveals that the parameter $|a|$ determines how rapidly the velocity changes with respect to the coordinate $x$, so it is a numerical measure of the quantity of shearing to which the plasma is subjected; this is consistent with its dual interpretation (see above) as the angular momentum parameter of the dual black hole, assuming that there is one. The combination $|a|/L^3$ is related to the value of the $x$ coordinate when $v$ reaches its maximum, so it is a measure of the transverse dimension of the interaction zone. Thus, all of the factors on the right side of equation (\ref{L}) are understood, and so \emph{we now have a physical interpretation for} $\ell$: it is a length scale determined by the transverse size, maximal velocity, and angular momentum of the shearing plasma.

Equation (\ref{K}) can be solved exactly, in terms of the Weierstrass $\wp$-function \cite{kn:abramo}. This function is characterised by a pair of numbers, $g_2$ and $g_3$, called the elliptic invariants, so that the function is expressed in general in the form $\wp\,(x\,;\,g_2,\,g_3)$. The solution is then
\begin{equation}\label{M}
v(x) = {V\over 3} - \wp\bigg({\sqrt{|a|}\over L^{3/2}}\big(x+\varepsilon\big);\; -4\bigg[1-{V^2\over 3}\bigg],\; {-8V \over 3}\bigg[1+{V^2\over 9}\bigg]\bigg).
\end{equation}
Here $\varepsilon$ is a constant chosen so that the graph passes through the origin.

Now bear in mind that the QGP exists only in the narrow strip $0 \leq x \leq X$, where $X$ is defined as half of the extent of the overlap of the two colliding nuclei; clearly $X$ can be computed from the transverse size of the nucleus and the impact parameter. Then $v(x)$ attains its maximum at the edge of this strip: that is, $v(X) = V$, or
\begin{equation}\label{MMMMM}
 {2V\over 3} + \wp\bigg({\sqrt{|a|}\over L^{3/2}}\big(X+\varepsilon\big);\; -4\bigg[1-{V^2\over 3}\bigg],\; {-8V \over 3}\bigg[1+{V^2\over 9}\bigg]\bigg) = 0.
\end{equation}
Combining this equation with $v(0) = 0$, we have a pair of simultaneous algebraic equations, from which the remaining parameters in equation (\ref{M}) can be computed, given $a$, $V$, and $X$; and so the shape of the velocity profile (in Figure 3, or the upper part of Figure 1) is fixed\footnote{In order to use the AdS/CFT correspondence in this manner of this work, one needs $L$ to be much larger than the string scale; this is the case here, since we find that it is consistent to choose $L$ to be of typical nuclear physics magnitude, of the order of 10 femtometres.}.

Note that the solution curve in this case has no point of inflection. One can see this from equation (\ref{K}) itself: differentiating and simplifying, we have
\begin{equation}\label{CONCAVEDOWN}
{\m{d}^2v\over \m{d}x^2}\;=\;-\,{2|a|\over L^3}\bigg[3\Big(v-{V\over 3}\Big)^2+\Big(1-{V^2\over 3}\Big)\bigg],
\end{equation}
which does not vanish. Therefore, once again, we are dealing with a classically stable fluid motion.

We see, then, that if we wish to use frame dragging at infinity to model the internal shearing motion of the QGP, then the boundary geometry is very tightly constrained even in the absence of detailed knowledge of the bulk geometry. In fact, since the function $v(x)$ is essentially the one non-trivial component of the boundary metric tensor in these coordinates, the boundary geometry is now completely determined, given physical data about the plasma and the region in which it forms. (Here we are leaving aside the problem of fitting Figures 2 and 3 together to obtain Figure 1; the complications arising from this will not affect our discussion, since our conclusions can be derived from considering each piece separately.)

In summary, the classification of possible ``holographic velocity profiles'' is complete: it is given above in terms of the classification of possible solutions of equation (\ref{I}). We have a holographic dual for profiles like those in Figure 2, and we can hope (see below) to find a dual for profiles of the type given in Figure 3. The holographic approach is therefore complementary to the classical methods used in \cite{kn:KelvinHelm,kn:viscous}, in the sense that it describes the situation everywhere in the fluid \emph{except} near to the zone where the Kelvin--Helmholtz instability is expected to arise.

It was argued in \cite{kn:75} that, when string-theoretic effects are taken into account, the KMV$_0$ black holes are unstable, and we will see below that the same is true of the QKMV$_0$ geometries, for all values of the charge. We need now to ask whether this instability afflicts black holes inducing at infinity profiles like the one in Figure 3; of course, that entails demonstrating that such black holes actually exist.

\addtocounter{section}{1}
\section* {\large{\textsf{3. The Generalised Planar Black Hole Geometry}}}
The QKMV$_0$ geometry discussed above was obtained \cite{kn:klemm} as a special case of the Pleba\'nski--Demia\'nski family of metrics \cite{kn:plebdem}. The latter has been expressed in a more convenient form by Griffiths and Podolsk\'y \cite{kn:grifpod,kn:exact}. It is this form that will be used here: see Appendix A for a brief review. Our objective is to find a member of this family which is asymptotically AdS and which induces a boundary geometry of the form (\ref{A}), with $\omega_{\infty}(x)$ such that equation (\ref{K}) is satisfied.

The boundary metric calculated from the Griffiths--Podolsk\'y metric (\ref{GP1a})--(\ref{GP1c}) is (replacing $p\rightarrow\psi$ and $\phi\rightarrow\zeta$ so as to agree with the QKMV$_0$ coordinates used above)
\ba
g({\rm GP})_\infty&=&-\dif t^2-2(\psi-1)(a\psi+a+2\ell)\,\dif t\dif\zeta\cr
&&+L^2\bigg[\frac{\dif\psi^2}{{\cal P}}+\Big[{\cal P}-\frac{1}{L^2}\big(a(1-\psi^2)+2\ell(1-\psi)\big)^2\Big]\,\dif\zeta^2\bigg].
\ea
Now we require the $g_{t\zeta}$ term to vanish at $\psi=0$. This is achieved by performing the shift $t\rightarrow t-(a+2\ell)\zeta$, resulting in the metric
\be
\label{boundary_metric}
g({\rm GP})_\infty=-\dif t^2-2\psi(a\psi+2\ell)\,\dif t\dif\zeta+L^2\bigg[\frac{\dif\psi^2}{{\cal P}}+{\cal R}\,\dif\zeta^2\bigg],
\ee
where
\be
{\cal R}\equiv a_0+a_1\psi-\Big(\epsilon-\frac{2\ell^2}{L^2}\Big)\psi^2,
\ee
and $a_0$, $a_1$ are given in (\ref{GP1c}) with $\varpi=1$. ${\cal P}$ and ${\cal R}$ should be positive to have the correct metric signature.

To obtain a planar metric, we require ${\cal R}$ to be a constant. This implies two conditions on the parameters $\epsilon$ and $n$:
\be
\label{epsn}
\epsilon=\frac{2\ell^2}{L^2}\,,\qquad n=0.
\ee
This last condition is especially interesting, because it actually expresses the condition that the boundary will be conformally flat. In other words, if a bulk metric of the form (\ref{GP1a})--(\ref{GP1c}) induces a boundary metric of the form given in equation (\ref{A}), then that metric will \emph{automatically} be conformally flat.

The remaining free parameter $k$ can then be chosen to set $a_0=1$. Together with the values (\ref{epsn}), we have
\be
{\cal R}=1\,,\qquad
{\cal P}=1+\frac{\psi^2}{L^2}(2\ell+a\psi)^2.
\ee
In particular, note that ${\cal P}$ is manifestly positive. The bulk metric (\ref{GP1a}) then becomes (replacing $p\rightarrow\psi$ and $\phi\rightarrow\zeta$)
\ba
\label{GP}
g(\ell {\rm QKMV}_0)&=&-\frac{\Delta_r}{\rho^2}\left[\dif
t-\big(a(1-\psi^2)+2\ell(1-\psi)\big)\,\dif\zeta\right]^2+\frac{\rho^2}{\Delta_r}\,\dif
r^2\nonumber\\
&&+\frac{\rho^2}{\Delta_\psi}\,\dif
\psi^2+\frac{\Delta_\psi}{\rho^2}\left[{a}\,\dif
t-\big(r^2+(\ell+a)^2\big)\,\dif\zeta\right]^2,
\ea
where
\ba
\Delta_r&=&\frac{(r^2+\ell^2)^2}{L^2}-8\pi M^*r+a^2+4\pi Q^{*2}\cr
\Delta_\psi&=&1+\frac{\psi^2}{L^2}(2\ell+a\psi)^2\cr
\rho^2&=&r^2+(\ell+a\psi)^2.
\ea
Here $M^*$ and $Q^*$ have slightly different physical interpretations from their QKMV$_0$ counterparts: for example, the charge density is $Q^*/(r^2_{\m{H}}+\ell^2)$ rather than $Q^*/r^2_{\m{H}}$, where $r_{\m{H}}$ as usual locates the event horizon.

If we perform the shift $t\rightarrow t+(a+2\ell)\zeta$, (\ref{GP}) can be written in the form
\be
\label{gen_QKMV}
g(\ell {\rm QKMV}_0)=-\frac{\Delta_r\Delta_\psi\rho^2}{\Sigma^2}\,\dif t^2+\frac{\rho^2\dif r^2}{\Delta_r}+\frac{\rho^2\dif\psi^2}{\Delta_\psi}+\frac{\Sigma^2}{\rho^2}\left[\omega\dif t-\dif\zeta\right]^2,
\ee
where
\ba
\label{Sigma}
\Sigma^2&=&(r^2+\ell^2)^2\Delta_\psi-\psi^2(a\psi+2\ell)^2\Delta_r\cr
\omega&=&\frac{\Delta_r\psi(a\psi+2\ell)+a(r^2+\ell^2)\Delta_\psi}{\Sigma^2}.
\ea
The corresponding electromagnetic potential is
\be
\label{AAAAA}
{\cal A}=-\frac{Q^*r}{\rho^2}\left[\dif
t+\psi(a\psi+2\ell)\dif\zeta\right].
\ee
It is easily checked that (\ref{gen_QKMV}) formally\footnote{The reason for the word ``formally'' can be seen in two ways: first, according to equation (\ref{L}), one cannot independently take $\ell$ and $a$ to zero ($V = 0$ does not make sense physically if $a \neq 0$); secondly, we saw that the parameter $a$ must be strictly positive in the QKMV$_0$ geometry, whereas any non-zero value of $\ell$ requires $a$ to be strictly negative if it does not vanish. Thus, the $\ell$QKMV$_0$ geometry cannot be regarded as a \emph{continuous} deformation of the QKMV$_0$ geometry.} reduces to the QKMV$_0$ metric (\ref{C}) if $\ell=0$: it represents a \emph{generalised planar black hole}, in the sense that it is specified not only by the mass and charge parameters, but also by the new parameter $\ell$. We shall call this the ``$\ell$QKMV$_0$ metric''. The meaning of the new parameter is not clear at this point, but it will be clarified by the holographic interpretation.

The positivity of $\Sigma^2$ can be checked by first rewriting it in the form
\be
\Sigma^2=r^4+2\ell^2r^2+8\pi\psi^2(a\psi+2\ell)^2M^*r+\cdots
\ee
where the ellipsis denotes terms independent of $r$. We immediately see that this is an increasing function of $r$ for positive $r$. Moreover, the expression of $\Sigma^2$ in (\ref{Sigma}) shows that it is positive on the horizon (defined by the largest root of $\Delta_r$), so it is positive everywhere outside the event horizon. It is clear that this geometry represents a well-defined planar black hole.

Note that the angular velocity of the event horizon is $\omega(r_{\rm H})=\frac{a}{r_{\rm H}^2+\ell^2}$, where $r = r_{\rm H}$ at the event horizon. It is independent of position there, but not at any other value of $r$, including infinity.

The special case $a = 0$, $\ell \neq 0$ merits attention, because of its great simplicity. Notice first that the metric is still non-diagonal in this case, so $\ell$ is more like angular momentum than electric charge in this case (see \cite{kn:chakra} for the corresponding statement in the spherical case). If we set $\dif\xi=\dif\psi/\sqrt{1+\frac{4\ell^2}{L^2}\psi^2}$, then  $\psi=\frac{L}{2\ell}\sinh\frac{2\ell\xi}{L}$. The boundary metric (\ref{boundary_metric}) then becomes
\be
\dif s^2_\infty=-\dif t^2-2L\sinh\frac{2\ell\xi}{L}\,\dif t\dif\zeta+L^2\big(\dif\xi^2+\dif\zeta^2\big).
\ee
This is precisely the boundary geometry we discovered in the preceding section in the special case $a = 0$, $\ell \neq 0$, which means that this particular boundary geometry \emph{does} have a dual interpretation in terms of a black hole in the bulk.

In fact, this statement holds true even when $a\neq 0$. To see this, we need to express the $\ell$QKNV$_0$ boundary metric explicitly in the form given in equation (\ref{A}). Notice first that in this case we have
\begin{equation}\label{N}
\omega_{\infty} = \psi (a\psi + 2\ell)/L^2.
\end{equation}
Next, define a new coordinate $\xi$ by
\begin{equation}\label{O}
\m{d}\xi^2 = {\m{d}\psi^2 \over 1+\frac{\psi^2}{L^2}(2\ell+a\psi)^2}.
\end{equation}
Setting $x=L\xi$, differentiating $\omega_{\infty}$, and squaring, we get
\begin{equation}\label{P}
\bigg({\m{d}\omega_{\infty}\over \m{d}x}\bigg)^2L^4 = 4\bigg({\m{d}\psi \over \m{d}\xi}\bigg)^2\Big({a\psi+\ell\over L}\Big)^2.
\end{equation}
A lengthy algebraic manipulation allows us to combine these equations to obtain
\begin{equation}\label{Q}
\bigg({\m{d}\omega_{\infty}\over \m{d}x}\bigg)^2L^4\;=\;4\bigg({\ell^2\over L^2}\,-\,{|a|\over L}\omega L\bigg)(1\,+\,\omega^2L^2).
\end{equation}
Defining, as usual, $v(x)$ by $v(x) = \omega_{\infty}L$, and $V$ as in equation (\ref{L}), we see that we have again obtained equation (\ref{K}) in the preceding section. There we obtained it from general properties of the boundary physics; here it arises from the bulk geometry. The Griffiths--Podolsk\'y parameter $\ell$ has turned out to be precisely the constant we obtained by solving the equation expressing the conformal flatness of the boundary. In other words, \emph{the geometry we discussed in the preceding section does have a bulk dual}: that dual is just the $\ell$QKMV$_0$ geometry. To put it yet another way: the boundary  metric of the form given in equation (\ref{A}), with $\omega_{\infty}(x)$ given (after dividing by $L$) in equation (\ref{M}), can be obtained by replacing the QKMV$_0$ geometry by an $\ell$QKMV$_0$ geometry, also in the Pleba\'nski--Demia\'nski class. The result is to replace Figure 2 by Figure 3 for the velocity profile at infinity.

As a side-benefit, $\ell$ acquires a clear physical interpretation in terms of the physics of the boundary field theory: as we saw earlier, it is a length scale computed in terms of the maximal dimensionless velocity of the plasma, its angular momentum per unit energy, and the AdS curvature scale. Notice however that causality in the boundary field theory, expressed as $V < 1$, imposes a bound on the ratio $\ell^2/(|a|L)$, namely
\begin{equation}\label{QQQQQ}
{\ell^2\over |a|L}\;<\;1.
\end{equation}
This looks rather restrictive from the bulk point of view, since the $\ell$QKMV$_0$ metrics violating it are classically well-defined. Later (Appendix B) we shall see, however, that there is good reason to impose this condition even in the bulk.

We see, then, that the $\ell$QKMV$_0$ geometries give rise to all of the conformally flat boundary geometries with velocity profiles as in Figure 3. This is at first sight surprising, because the Pleba\'nski--Demia\'nski geometries we have considered are by no means the most general ones: we have assumed, merely for simplicity, that the ``acceleration'' parameter $\alpha$ (see Appendix A) vanishes. Since we have just seen that the $\alpha = 0$ bulk geometries already account for all possible boundary geometries of the kind we need, it follows that taking $\alpha \neq 0$ cannot lead to anything new; the $\ell$QKMV$_0$ geometries are the most general ones of Pleba\'nski--Demia\'nski type inducing boundary geometries representing the QGP generated by a peripheral heavy-ion collision.

In order to complete the holographic dictionary in this case, we need to show how the bulk parameters $M^*$ and $Q^*$ can be computed from boundary data. Here the \emph{temperature} and \emph{chemical potential} of the boundary field theory are relevant. To compute them, it is as usual convenient to transform to the Euclidean domain, which we now explore.

\addtocounter{section}{1}
\section* {\large{\textsf{4. Temperature and Chemical Potential}}}
We now discuss the Euclidean version of the $\ell$QKMV$_0$ geometry. We will need this to compute the temperature of the black hole, but it is also needed for studying the chemical potential of the boundary field theory, and to compute the action for certain branes propagating in the bulk. Technically, our focus is on the periodicities of the Euclidean versions of the non-radial coordinates (including time).

\subsubsection*{{\textsf{4.1 Euclidean $\boldsymbol\psi$}}}
To understand the Euclidean version of the coordinate $\psi$, it is best to consider the situation on the boundary. The Euclidean version of equation (\ref{K}) is
\begin{equation}\label{R}
\bigg({\m{d}v_{\rm E}\over \m{d}x}\bigg)^2\;=\;{4|a|\over L^3}\big(V - v_{\rm E}\big)\big(1\,-\,v_{\rm E}^2\big);
\end{equation}
note that $v$ and $V$, being time derivatives, have to be complexified (in the same way as $|a|$ and $\ell$, that is, $|a| \rightarrow -i|a|$, $\ell \rightarrow -i\ell$, $V \rightarrow -iV$, $v \rightarrow -iv_{\rm E}$). This remarkable differential equation has two disjoint sets of solutions. One set consists of unbounded functions, bounded below by unity; these are the functions one obtains by complexifying the parameters in the Weierstrass $\wp$-function in equation (\ref{M}). These are of no interest here, since they will not define a complete Euclidean metric. The other set consists of functions satisfying $-1 \leq v_{\rm E} \leq V$. The solution in this case takes the form
\be
v_{\rm E}=\frac{(1+V)(1+\sqrt{2-2V})\,{\rm sn}(Cx+D,k)-(1-V)\sqrt{2-2V}+3V-1}{(1+V)\,{\rm sn}(Cx+D,k)+2\sqrt{2-2V}-V+3},
\ee
where $D$ is arbitrary and sn$(x,k)$ is, as mentioned above, one of the Jacobi elliptic functions \cite{kn:abramo}; here
\be
C=\sqrt{\frac{2|a|}{L^3}}\,\frac{2\sqrt{2-2V}-V+3}{2+\sqrt{2-2V}},\qquad    k=\frac{(1+V)}{2\sqrt{2-2V}-V+3}.
\ee
This solution was obtained by performing a fractional linear transformation in $v_{\rm E}$ to turn (\ref{R}) into a well-known differential equation whose solution is just sn$(x,k)$.

Now sn$(x,k)$ is periodic on the real axis, hence the same is true of $v_{\rm E}$: in other words, the coordinate $x$ is compactified in the Euclidean case. Its period can be computed from the period of sn$(x,k)$, usually stated in terms of the complete elliptic integrals of the first kind \cite{kn:abramo}. However, the periodicity can be stated more usefully in terms of the period of the Euclidean version of the coordinate $\psi$. (Note from the Euclidean version of equation (\ref{O}) that $\psi$ can be expressed in terms of $\xi$, to which $x$ is proportional.) Using the Euclidean version of equation (\ref{N}), one sees that, as $v_{\rm E}$ passes from one minimum ($v_{\rm E} = -1$) to the next, $\psi$ varies between
$\Psi_1$ and $\Psi_2$, where
\ba
\label{Psi}
\Psi_1&=&\frac{-\ell+\sqrt{\ell^2-aL}}{a}\cr
\Psi_2&=&\frac{-\ell-\sqrt{\ell^2-aL}}{a}.
\ea
These quantities\footnote{Note that, because the form of equation (\ref{N}) is unchanged by complexification (there being a common factor of $-i$ throughout), these expressions are invariant under complexification. Notice too that the geometry is symmetrical not about $\psi =0$, but rather about $\psi = -\ell /a$, that being the arithmetic mean of $\Psi_1$ and $\Psi_2$. As might be expected, this statement is true also in the Lorentzian case: see below.} may therefore be taken as defining the range of $\psi$, on the boundary and therefore in the bulk.

\subsubsection*{{\textsf{4.2 Euclidean $\boldsymbol t$ and $\boldsymbol\zeta$}}}
The Euclidean versions of the coordinates $t$ and $\zeta$ are best studied in the bulk.

The Euclidean version of the bulk metric (\ref{gen_QKMV}) is
\be
\label{EUCLID}
g({\rm E\ell QKMV}_0)=\frac{\Delta_r^{\rm E}\Delta_\psi^{\rm E}\rho^2_{\rm E}}{\Sigma^2_{\rm E}}\,\dif t^2+\frac{\rho^2_{\rm E}\dif r^2}{\Delta_r^{\rm E}}+\frac{\rho^2_{\rm E}\dif\psi^2}{\Delta_\psi^{\rm E}}+\frac{\Sigma^2_{\rm E}}{\rho^2_{\rm E}}\left[\omega_{\rm E}\dif t-\dif\zeta\right]^2,
\ee
where
\ba
\label{EUCLIDMORE}
\Delta_r^{\rm E}&=&\frac{(r^2-\ell^2)^2}{L^2}-8\pi M^*r-a^2-4\pi Q^{*2}\cr
\Delta_\psi^{\rm E}&=&1-\frac{\psi^2}{L^2}(2\ell+a\psi)^2\cr
\rho^2_{\rm E}&=&r^2-(\ell+a\psi)^2\cr
\Sigma^2_{\rm E}&=&(r^2-\ell^2)^2\Delta_\psi^{\rm E}+\psi^2(a\psi+2\ell)^2\Delta_r^{\rm E}\cr
\omega_{\rm E}&=&\frac{\Delta_r^{\rm E}\psi(a\psi+2\ell)+a(r^2-\ell^2)\Delta_\psi^{\rm E}}{\Sigma^2_{\rm E}}.
\ea
We may call this the ``$\ell$QKMV$_0$ gravitational instanton''. The corresponding Euclidean electromagnetic potential is
\be
\label{BBBBB}
{\cal A}^{\rm E}=-\frac{Q^*r}{\rho_{\rm E}^2}\left[\dif
t-\psi(a\psi+2\ell)\dif\zeta\right].
\ee

If $r_{\rm H}^{\rm E}$ denotes the value of the radial coordinate at the Euclidean event horizon, then
\be
8\pi M^*=\frac{((r_{\rm H}^{\rm E})^2-\ell^2)^2-(a^2+4\pi Q^{*2})L^2}{r_{\rm H}^{\rm E}L^2},
\ee
and $\Delta_r^{\rm E}$ can be written in the alternative, factorised form:
\be
\label{Delta_r}
\Delta_r^{\rm E}=(r-r_{\rm H}^{\rm E})\,\frac{rr_{\rm H}^{\rm E}(r+r_{\rm H}^{\rm E})^2-(rr_{\rm H}^{\rm E}+\ell^2)^2+(a^2+4\pi Q^{*2})L^2}{r_{\rm H}^{\rm E}L^2}.
\ee

To ensure a positive definite signature, we restrict ourselves to the range $r\geq r_{\rm H}^{\rm E}>\sqrt{\ell^2+|aL|}>0$ and $\Psi_1\leq\psi\leq\Psi_2$, where $\Psi_{1,2}$ are roots of $\Delta_\psi^{\rm E}$, appropriately chosen from the four possibilities $\frac{-\ell\pm\sqrt{\ell^2\pm aL}}{a}$. (Note that the quantities we previously called $\Psi_1$ and $\Psi_2$ correspond to a particular pair chosen from these four.) In particular, these ranges will ensure that $\rho_{\rm E}^2$ and the second factor of (\ref{Delta_r}) are positive.

Recall that we need $a < 0$. There are two separate cases to consider: $\ell^2/(|a|L) < 1$ and $\ell^2/(|a|L) > 1$. The first case is the physical one here: the second would violate causality on the boundary, as we discussed earlier (inequality (\ref{QQQQQ})). It turns out (see Appendix B) that the second case is also unacceptable from the bulk point of view, because it does not correspond to a well-defined gravitational instanton. We therefore confine ourselves here to the first case. In this case, the relevant (real) roots of $\Delta_\psi^{\rm E}$ are precisely those given in equations (\ref{Psi}) above.

Now since the $\ell$QKMV$_0$ gravitational instanton has two commuting Killing vector fields $\partial_t$ and $\partial_{\zeta}$, with their associated fixed points at $r=r_{\rm H}^{\rm}$ and $\psi=\Psi_{1,2}$, it has a $U(1)\times U(1)$ isometry group. A well-developed formalism (see \cite{kn:Chen}) of ``rod structures'' has been constructed to deal with such gravitational instantons, and we will use it here. Briefly, this formalism assigns a sequence of so-called rods to a given gravitational instanton, with each rod having a certain direction $a\partial_t+b\partial_\zeta$. Each rod actually represents a two-dimensional fixed point set of the $U(1)$ isometry subgroup generated by its direction. It follows that points at which two adjacent rods meet are fixed points of the whole isometry group.

It can be checked that the rod structure of the $\ell$QKMV$_0$ gravitational instanton consists of the following three rods:
\begin{itemize}
\item Rod 1: a semi-infinite rod located at $\psi=\Psi_{1}$, with direction
\be
\label{l1}
\varrho_1=\frac{L}{2\sqrt{\ell^2-aL}}\left(-L\partial_t+\partial_\zeta\right).
\ee
\item Rod 2: a finite rod located at $r=r_{\rm H}^{\rm E}$, with direction
\be
\label{l2}
\varrho_2=\frac{2\big((r_{\rm H}^{\rm E})^2-\ell^2\big)}{{\Delta^{\rm E}_{r}}'(r_{\rm H}^{\rm E})}\left({\partial_t}+\frac{a}{(r_{\rm H}^{\rm E})^2-\ell^2}\,{\partial_\zeta}\right).
\ee
\item Rod 3:  a semi-infinite rod located at $\psi=\Psi_{2}$, with the same direction as Rod 1: $\varrho_{3}=\varrho_1$.
\end{itemize}
The pair $\{\varrho_1,\varrho_2\}$, or equivalently the pair $\{\varrho_2,\varrho_3\}$, generates the $U(1)\times U(1)$ isometry of this gravitational instanton.

Let us quickly work out the explicit identifications imposed on the coordinates $(t,\zeta)$ by this isometry, and the requirement that the gravitational instanton is regular. If we change coordinates
\ba
t&=&t'-L\zeta'\cr
\zeta&=&\frac{a}{(r_{\rm H}^{\rm E})^2-\ell^2}\,t'+\zeta',
\ea
the two linearly independent rod directions take the ``orthogonal'' form:
\be
\varrho_1=\frac{L}{2\sqrt{\ell^2-aL}}\,{\partial_{\zeta'}},
\qquad
\varrho_2=\frac{2\big((r_{\rm H}^{\rm E})^2-\ell^2\big)}{{\Delta^{\rm E}_{r}}'(r_{\rm H}^{\rm E})}\,{\partial_{t'}}.
\ee
In order to avoid conical singularities along the rods, we then have to impose the following two independent identifications:
\ba
t'&\rightarrow&t'+\frac{4\pi\big((r_{\rm H}^{\rm E})^2-\ell^2\big)}{{\Delta^{\rm E}_{r}}'(r_{\rm H}^{\rm E})}\cr
\zeta'&\rightarrow&\zeta'+\frac{\pi L}{\sqrt{\ell^2-aL}}.
\ea
In terms of the original coordinates $(t,\zeta)$, these two identifications are
\ba
\label{IDENT}
(t,\zeta)&\rightarrow&\bigg(t+\frac{4\pi\big((r_{\rm H}^{\rm E})^2-\ell^2\big)}{{\Delta^{\rm E}_{r}}'(r_{\rm H}^{\rm E})},\zeta+\frac{4\pi a}{{\Delta^{\rm E}_{r}}'(r_{\rm H}^{\rm E})}\bigg)\\
\label{IDENT2}
(t,\zeta)&\rightarrow&\bigg(t-\frac{\pi L^2}{\sqrt{\ell^2-aL}},\zeta+\frac{\pi L}{\sqrt{\ell^2-aL}}\bigg),
\ea
both of which non-trivially mix the two coordinates.

It can be seen from the above rod structure that constant $r$ and $t'$ sections of the gravitational instanton are topological two-spheres parameterised by $(\psi,\zeta')$, with the two ``poles'' at $\psi=\Psi_{1,2}$. It follows that the constant $r$ sections have $S^2\times S^1$ topology. In particular, the boundary of the $\ell$QKMV$_0$ gravitational instanton inherits this compact topology.

\subsubsection*{{\textsf{4.3 Temperature of the $\boldsymbol\ell$QKMV$_0$ Planar Black Hole}}}
As usual, the temperature of the black hole Hawking radiation (see \cite{kn:jesus}) can now be computed from the periodicity of Euclidean ``time'': we find, after continuing back to the Lorentzian section,
\begin{equation}\label{S}
T \;=\;{r_{\rm H}(r_{\rm H}^2+\ell^2)/L^2-2\pi M^* \over \pi (r_{\rm H}^2+\ell^2)}.
\end{equation}
Recall that $\ell$ can be expressed in terms of the parameters $V$, $|a|$, and $L$, all of which can be computed from field theory data on the boundary, and that $r_{\rm H}$ is related to the other parameters by the equation $\Delta_r(r_{\rm H}) = 0$; in the zero-charge case we can now compute $M^*$ given the temperature. In the charged case, however, we need another equation, relating the charge parameter $Q^*$ to physical data. We now turn to this.

\subsubsection*{{\textsf{4.4 Relating the Charge Parameter to the Field Theory Chemical Potential}}}
The Euclidean electromagnetic potential form in the $\ell$QKMV$_0$ geometry was given earlier (equation (\ref{BBBBB})). It can be expressed more generally as
\begin{equation}\label{T}
{\cal A}^{\rm E} = \Big[-{Q^*r \over  \rho_{\rm E}^2}+\kappa_t^{\rm E}\Big]\dif t + \Big[{Q^*r \over  \rho_{\rm E}^2}\psi(a\psi + 2\ell)+\kappa_{\zeta}^{\rm E}\Big]\dif\zeta,
\end{equation}
where $\kappa_t^{\rm E}$ and $\kappa_{\zeta}^{\rm E}$ are constants corresponding to a choice of gauge; notice that they also give the asymptotic value of ${\cal A}^{\rm E}$. Now recall from the rod structure formalism that the norms of both $\partial_t$ and $\partial_{\zeta}$ vanish at those points where the rods intersect; that is, when $r = r_{\rm H}^{\rm E}$ and $\psi = \Psi_1$ (or $\Psi_2$; see equations (\ref{Psi})). Since the geometry is indeed Euclidean, this means that the vector fields $\partial_t$ and $\partial_{\zeta}$ themselves vanish at these points, which in turn means that ${\cal A}^{\rm E}(\partial_t) = {\cal A}^{\rm E}(\partial_{\zeta}) = 0$ there; hence we see that, if $\rho^{\#}_{\rm E}$ is the value of $\rho_{\rm E}$ at $r = r_{\rm H}^{\rm E}$ and $\psi = \Psi_1$, then

\begin{equation}\label{U}
\kappa_t^{\rm E} = {Q^*r^{\rm E}_{\rm H} \over (\rho_{\rm E}^{\#})^2}
\end{equation}
and
\begin{equation}\label{V}
\kappa_{\zeta}^{\rm E} = -{Q^*r^{\rm E}_{\rm H} \over (\rho_{\rm E}^{\#})^2}\Psi_1(a\Psi_1+2\ell).
\end{equation}

After some simplifications followed by complexification one can obtain the Lorentzian versions of these quantities, and so evaluate the asymptotic electromagnetic potential form:
\begin{equation}\label{W}
{\cal A}_{\infty} = {Q^*r_{\rm H} \over r_{\rm H}^2+\ell^2+|a|L}[\dif t - \dif z].
\end{equation}
This can be used to compute the electric potential with respect to any given choice of observer at infinity. We are particularly interested in the observer who is at rest relative to the axis of the collision, whose worldline has unit (with respect to the boundary metric) tangent vector $\partial_t$; in the AdS/CFT dictionary, the quark chemical potential $\mu$ is proportional \cite{kn:koba} to the asymptotic electric potential seen by this observer. Evaluating ${\cal A}_{\infty}$ on $\partial_t$, and regularizing the units according to \cite{kn:myers}, we obtain finally
\begin{equation}\label{X}
\mu = {Q^*r_{\rm H} \over \pi L[r_{\rm H}^2+\ell^2+|a|L]}.
\end{equation}
Given the temperature, chemical potential, and angular momentum per unit energy of the boundary theory, we now have three equations (equation (\ref{S}), $\Delta_r(r_{\rm H}) = 0$, and equation (\ref{X})) for the three unknowns $M^*$, $Q^*$, and $r_{\rm H}$. In short, the bulk geometry is now fully computable from the boundary data.

\addtocounter{section}{1}
\section* {\large{\textsf{5. Branes in the $\boldsymbol\ell$QKMV$_0$ Geometry}}}
According to the AdS/CFT duality, string theory on the background defined by the black hole we have been discussing is equivalent to the physics of a certain field theory on a specific conformally Minkowskian geometry, of a kind which can be used to model a system undergoing shear. That field theory is not QCD, but one can hope, with due caution \cite{kn:mateos,kn:karch,kn:kovch}, that the differences are not so great as to render the comparison valueless. In short, the behaviour of string theory on the $\ell$QKMV$_0$ background might have something new to teach us about the shearing QGP.

One of the most characteristic properties of (Euclidean) asymptotically AdS spaces is that the areas of surfaces are comparable to the volumes they enclose, as they expand towards infinity. This becomes a \emph{directly physical} effect in string theory, because of the presence of branes wrapping around such surfaces.

The action of a brane has a positive contribution from its tension (related to the area) but also a \emph{negative} contribution from the enclosed volume (due to the coupling to the background). This negative term does not lead to any pathologies in AdS itself, but Seiberg and Witten \cite{kn:seiberg} pointed out that it might do so, for BPS branes, in spaces which are only \emph{asymptotically} AdS: the geometric deformation might upset the delicate competition between the area and volume terms, allowing the latter to dominate, leading perhaps to a brane action which is unbounded below. If that should happen, the resulting pair-production instability would correspond, through the AdS/CFT correspondence, to the existence of some kind of instability in the field theory. Because of its generality, this is a very important effect, and it has been used to address some fundamental issues in AdS/CFT theory (for example, see \cite{kn:wittenyau}).

This issue was considered in the case of the KMV$_0$ geometry in \cite{kn:75}. We found there that this effect does arise in that case. The question is whether this is also true of the QKMV$_0$ and $\ell$QKMV$_0$ geometries, and, if so, for which values of the new parameters.

In the case of the Euclidean version of the $\ell$QKMV$_0$ geometry, we can compute the Seiberg--Witten action for a brane wrapping a compact transverse section at any fixed value of the radial coordinate $r$: it takes the form
\be
{\rm E}\SW=\Theta^{\rm E} \bigg[\int_{\Psi_1}^{\Psi_2}\sqrt{\Delta_r^{\rm E}}\rho_{\rm E}\,\dif \psi-\frac{3}{L}\int_{r_{\rm H}^{\rm E}}^r\int_{\Psi_1}^{\Psi_2}\rho_{\rm E}^2\,\dif \psi\dif r\bigg].
\ee
Here $\Theta^{\rm E}$ is a constant whose precise value need not detain us\footnote{$\Theta^{\rm E}$ is a combination of the tension per unit area of the brane with the volume of the compact space in the $t$, $\zeta$ directions. This volume can be computed from the identifications in equations (\ref{IDENT}), (\ref{IDENT2}); it does not depend on $r$. In this work we are only concerned with the zeros of the action function, and so the precise value of this constant, and of its Lorentzian version, do not affect our discussion.}. The first term corresponds to the area of the brane, computed in the Euclidean geometry described by the metric in equation (\ref{EUCLID}), while the second corresponds to the volume enclosed by that brane, measured from the origin of the Euclidean space (at $r = r_{\rm H}^{\rm E}$).

Evaluating the integrals, we have
\be
\int_{r_{\rm H}^{\rm E}}^r\int_{\Psi_1}^{\Psi_2}\rho_{\rm E}^2\,\dif \psi\dif r=-\frac{2}{3a}\sqrt{\ell^2-aL}\,\Big[r^3-(r_{\rm H}^{\rm E})^3-(r-r_{\rm H}^{\rm E})(\ell^2-aL)\Big],
\ee
\be
\int_{\Psi_1}^{\Psi_2}\sqrt{\Delta_r^{\rm E}}\rho_{\rm E}\,\dif \psi=-\frac{1}{a}\sqrt{\Delta_r^{\rm E}}\left(\sqrt{\ell^2-aL}\sqrt{r^2-\ell^2+aL}+r^2\arcsin\frac{\sqrt{\ell^2-aL}}{r}\right).
\ee
After an analytic continuation back to Lorentzian signature, the physical brane action is found to be\footnote{The action for branes in the QKMV$_0$ geometry may be obtained from this formula by formally setting $\ell = 0$ and reversing the sign of $a$.}
\ba
\label{ACTION}
\SW(r)&=&\Theta\, \bigg\{-\frac{1}{a}\sqrt{\Delta_r}\left(\sqrt{\ell^2-aL}\sqrt{r^2+\ell^2-aL}+r^2{\rm arcsinh}\,\frac{\sqrt{\ell^2-aL}}{r}\right)\cr
&&\hskip1.5cm +\frac{2}{aL}\sqrt{\ell^2-aL}\,\Big[r^3-r_{\rm H}^3+(r-r_{\rm H})(\ell^2-aL)\Big]
\bigg\}.
\ea

By construction, this action vanishes at the event horizon, and it is positive at first as $r$ increases. \emph{But it does not remain positive}: eventually the graph of the action as a function of $r$ turns over and cuts the horizontal axis. In fact, at large $r$, $\SW(r)$ behaves as
\ba
\SW(r)&\simeq& C\;+\Theta \bigg[-\frac{(\ell^2+5aL)\sqrt{\ell^2-aL}}{3aL}\bigg]\,r\cr
&=&C\;-\,\Theta\bigg[\frac{(5-V)\sqrt{(V+1)|a|L}}{3}\bigg]\,r,
\ea
where $C$ is a certain positive constant. Since $V$ is less than unity, we see that the graph at large $r$ is essentially that of a straight line with negative slope, so it will ultimately cut the horizontal axis and remain negative. In short, the brane action is unbounded below at large values of $r$, and the system is \emph{unstable} to brane pair-production \cite{kn:maoz} for all non-zero amounts of shear, in all of the QKMV$_0$ and $\ell$QKMV$_0$ geometries\footnote{Notice that, if $a$ does not vanish, then the system is unstable for \emph{all} values of $Q^*$: one can show that this is not so when $a$ and $\ell$ are exactly zero. One can also show, by a separate calculation, that the instability is present, for all values of $\ell$ and $Q^*$, in the case with $a = 0, \ell \neq 0$.}. (The result follows in the former case by formally setting $V = 0$ and reversing the sign of $a$.) By holography, it follows that the shearing boundary theory is likewise unstable, in every case.

The final conclusion is that, according to holography, \emph{both of the flows portrayed in Figures 2 and 3 are unstable}. It follows that a realistic flow for the QGP in the aftermath of a heavy-ion collision, such as the one shown in Figure 1, is unstable not just in the centre, where there is certainly a Kelvin--Helmholtz instability, but throughout the flow field. In short, holography suggests that classical hydrodynamics may be misleading us here, in a somewhat dramatic manner.

However, such plasmas exist, in any case, only for a very short period of time (of the order of 5 fm/c, though somewhat longer in the most energetic LHC collisions). The simulations in \cite{kn:KelvinHelm} indicate that classical instabilities in the QGP flow develop extremely rapidly, and it is clear that they must do so if the effect is to be physically relevant. If the instability we have been discussing does not likewise evolve very rapidly, \emph{then it will not influence the results of any experiment}. Thus, we cannot draw any conclusions until we have some way of estimating, even very roughly, the time scale for the development of Seiberg--Witten instability. We suggest that holography itself provides such an estimate, as we now explain.

\addtocounter{section}{1}
\section* {\large{\textsf{6. A Holographic Estimate of the Time Scale}}}
Using the holographic duality to estimate time scales in the evolution of strongly coupled systems is a well-established procedure; see for example \cite{kn:chesler} and references therein. In this spirit, let us try to estimate time scales for the hydrodynamic instability we discussed in the preceding section, by studying the evolution of the bulk instability we identified there.

We saw that the problem arises at \emph{large} values of $r$: that is to say, not in the immediate vicinity of the black hole. One can picture the situation as follows: the black hole is surrounded by a ``screen'', with a location determined by the value of the radial coordinate where the brane action cuts the horizontal axis: the system only misbehaves, initially, outside this screen.

Now in AdS/CFT duality, the boundary theory is dual to (at least) the \emph{entire} bulk system outside the event horizon. The suggestion is that the time required for the boundary instability to dominate should correspond to the time required for the region outside the black hole but inside the screen to be overwhelmed by the dynamics of the region of negative action. This in turn can be crudely estimated by asking how long it takes for a particle to fall through the screen to the event horizon. The region of negative action is defined by $r > r_{\rm S}$, where $r_{\rm S}$ is the value of $r$ at which the action function vanishes: this is the location of the ``screen''. Thus we are asking for the proper time required for a particle to fall from $r = r_{\rm S}$ to $r = r_{\rm H}$, the location of the event horizon.

There is a technical complication here, arising from the fact that, like any black hole with non-zero angular momentum, the $\ell$QKMV$_0$ black holes are not isotropic: that is, just as the ordinary Kerr metric is not symmetrical in the latitudinal direction, the $\ell$QKMV$_0$ black hole is not symmetrical in the $\psi$ direction. What this means is that, in general, a geodesic beginning at a point with a given value of the coordinate $\psi$ will not maintain that value as the geodesic is extended towards the event horizon. At the boundary, this means (see equation (\ref{N})) that the corresponding observer's velocity changes with time; that is, we would be computing times relative to a non-inertial observer. To solve this problem, we must begin with a value of $\psi$ which does not change, so that we are dealing with an inertial boundary observer. The times thus found can be converted to the laboratory frame by means of a Lorentz transformation.

If we define $\varphi \equiv \ell + a\psi$, then one can readily verify that the $\ell$QKMV$_0$ geometry is symmetrical in $\varphi$, in the sense of being invariant under $\varphi \rightarrow - \varphi$; in other words, the geometry is symmetrical about $\psi = -\ell/a = \ell/|a|$. It follows that the hypersurface $\psi = \ell/|a|$ is totally geodesic ---$\,$ a geodesic initially tangential to that surface will remain in it\footnote{The geometric significance of $\psi = \ell/|a|$ is as follows: if one computes the proper distance from an event $(t_0, r_{\rm S}, \psi, \zeta_0)$ on the screen to a simultaneous event $(t_0, r_{\rm H}, \psi, \zeta_0)$ on the event horizon, and allows $\psi$ to vary, then the distance will be smallest when $\psi = \ell/|a|$. That is, we are considering the direction in which the screen is nearest to the event horizon. (Bear in mind however that an object released from $(t_0, r_{\rm S}, \psi, \zeta_0)$ will not reach the event horizon at a point with $\zeta = \zeta_0$, because of frame-dragging.)}. According to equation (\ref{N}), this corresponds to an inertial observer on the boundary who is moving at a \emph{constant} dimensionless velocity $V$ ---$\,$ that is, at the maximal velocity of the plasma in the laboratory frame. In order to obtain times measured in the laboratory frame itself, then, we need to compensate with a factor of $\sqrt{1 - V^2}$.

With all this in mind, let us proceed. We take it that the object has zero momentum initially. It is initially at rest in the radial direction, and $\psi = \ell/|a|$ permanently, as above; due to frame dragging, however, zero momentum in the $\zeta$ direction does not imply that $\zeta$ is a constant. The existence of the two Killing vector fields, $\partial_t$ and $\partial_{\zeta}$, gives us two constants of the motion ($K$ in the first case, zero in the second, by assumption):
\begin{equation}\label{Y}
{- \Delta_r \;+\;a^2\Delta_{\psi} \over r^2}\,\dot{t} - {\Sigma^2\over r^2}\,\omega\,\dot{\zeta} = K
\end{equation}
\begin{equation}\label{Z}
- \omega \,\dot{t} + \dot{\zeta} = 0.
\end{equation}
Here the dot denotes a derivative with respect to proper time. The fact that the tangent of the geodesic, parametrised by proper time, is a unit vector, gives us a third equation. These can be combined to obtain $\dot{t}$, $\dot{\zeta}$, and $\dot{r}$. The solution for $\dot{r}$ gives us, after some simplifications and with the addition of the Lorentz factor, the time we seek:
\begin{equation}\label{ALPHA}
\tau \;=\;\sqrt{1 - V^2}\,\int_{r_{\rm H}}^{r_{\rm S}} \bigg[{-\Delta_r\over r^2}-K^2\bigg({V^2L^2\Delta_r\over r^4(1+V^2)}-\Big(1+{V|a|L\over r^2}\Big)^2\bigg)\bigg]^{-\,1/2}\m{d}r.
\end{equation}
Here $V$ has its usual meaning, and $K$ is to be fixed by the requirement that $\dot{r}$ should vanish at $r = r_{\rm S}$. The expression $\Delta_r$ depends on the parameters $M^*$ and $Q^*$, but we now know how to fix these, as well as $r_{\rm H}$, in terms of boundary data. Finally, $r_{\rm S}$ can likewise be computed from those data, by setting the expression in equation (\ref{ACTION}) equal to zero. Thus $\tau$ can be determined, given physical information regarding the boundary theory.

Let us try to implement this idea with actual estimates of the relevant numbers. Even if, as seems very probable, the results are too rough to provide anything more than a qualitative impression of the time scale itself, they might yield hints as to whether it roughly of a magnitude such that the instability can affect the plasma before it hadronises; furthermore, it seems likely that they could give a useful indication of \emph{trends} in the time scale as the parameters of the black hole are modified ---$\,$ that is, by duality, as the physical parameters of the boundary theory are varied. Such variations are in fact the subject of current and near-future experimental programmes (``beam energy scans'' of the quark matter phase diagram).

In \cite{kn:75}, we used data from the RHIC experiment \cite{kn:phobos} to estimate a \emph{maximal} possible value of the angular momentum density in the relevant collisions of $\approx$ 360/fm$^3$; maximal in the sense that the impact parameter was chosen so that the angular momentum transfer to the plasma should be as large as possible (see \cite{kn:bec}). Since a generic collision will give rise to somewhat lower angular momenta, and because the beam energy scan experiments involve less energetic collisions, we use 15 fm (based on an energy density estimate of $\approx$ 3 GeV/fm$^3$) as a conservative estimate for $a$ in a typical collision. Assuming that the impact parameters of the most relevant collisions here have values centred on the optimal value given in \cite{kn:bec}, one can compute the diameter of the corresponding interaction zone for gold nuclei; half of this is the range of the coordinate $x$ in Figures 1-3, and a reasonable estimate is about 6 fm. The full ranges of the temperatures and chemical potential values to be explored in the beam energy scan experiments are not known precisely, but it is safe to assume that they will be contained in the region $0.5 < T < 2$, $0 < \mu < 5$, both in units of fm$^{-1}$.

Finally, we need to estimate $V$, the maximal velocity of the plasma. This is more difficult, because there may be considerable deceleration during the brief interval after the collision but before the hydrodynamic regime begins. Following \cite{kn:KelvinHelm}, we take $V$ $\approx$ 0.4; the associated value of $L$ is $\approx$ 11.35 fm. The results are mildly sensitive\footnote{For example, if $V$ were $\approx$ 0.5 (which is by no means unreasonable) then, for $T$ = 0.5 fm$^{-1}$, $\mu$ = 2 fm$^{-1}$, $\tau$ would drop from 6.9739 fm/c to 6.1727 fm/c.} to the choice of $V$; clearly a better understanding of this parameter would be very desirable.

With these parameter choices, which have been made mainly for illustrative purposes, we have computed the relevant values of $\tau$ for both the QKMV$_0$ and the $\ell$QKMV$_0$ geometries (that is, for flows of the forms given by Figures 2 and 3 respectively). We find that the results do not differ very greatly in the two cases; they are shorter in the $\ell$QKMV$_0$ case, by roughly 10\%\footnote{For example, for $T$ = 0.5 fm$^{-1}$, $\mu$ = 2 fm$^{-1}$, $\tau$ is 7.6848 fm/c for the QKMV$_0$ case, 6.9739 fm/c for the $\ell$QKMV$_0$ geometry.}, which should not be regarded as meaningful in this context. We report the numerical data only for the $\ell$QKMV$_0$ case.

The results are shown in the table: times are in units of fm (usually denoted by fm/c).
\begin{center}
\begin{tabular}{|c|c|c|c|c|c|c|}
  \hline
 & $\mu$ = 0  &  $\mu$ = 1 & $\mu$ = 2 &$\mu$ = 3 &$\mu$ = 4&$\mu$ = 5 \\
\hline
$T$ = 2 &    16.3377  & 16.3299 & 11.3672 & 9.2306 & 8.2665 & 7.7283\\
$T$ = 1 &  16.3305 & 11.3642 & 8.2661 & 7.3857 & 6.9741 & 6.7348\\
$T$ = 0.5 &  16.3016  & 8.2649  & 6.9739 & 6.5780 & 6.3846 &6.2698
 \\
\hline
\end{tabular}
\end{center}
Recall that the QGP is expected to survive for a time of the order of 5 fm/c, though somewhat more in very energetic collisions. Again, we stress that the numerical values should be trusted only on the qualitative level: the trends, however, are more reliable. The most striking trend is of course the remarkable decrease in $\tau$ as $\mu$ increases and $T$ decreases, as will happen in the beam energy scans \cite{kn:ilya,kn:dong,kn:shine,kn:fair,kn:nica}. Intuitively, the screen around the black hole in the bulk is ``thinned'' by larger values of its electric charge, particularly at low Hawking temperatures; the dual of this phenomenon produces the pattern in the table.

A reasonably conservative interpretation of these results is the following. The model suggests that, contrary to expectations based on classical hydrodynamics, the flow of the QGP is unstable throughout the fluid. However, it may be that this non-classical instability arises too slowly to be important when the quark chemical potential is very small ---$\,$ as it has been in most observed collisions to date. The situation is less clear, however, for the larger values of $\mu$ being investigated in current beam energy scan programmes, or in those of the near future, since the time required drops rather dramatically under those conditions. Our tentative conclusion is that, as assumed in \cite{kn:KelvinHelm}, the plasma flow is \emph{effectively} stable, away from the point of inflection in the velocity profile, at low values of $\mu$; but that it is not safe to assume this in future experiments involving very high values of $\mu$.

\addtocounter{section}{1}
\section* {\large{\textsf{7. Conclusion}}}
In this work we have exhibited planar black holes of the Pleba\'nski--Demia\'nski form, and we have shown how they may be physically interpreted in the context of the holographic description of strongly coupled fluids. We have found that these bulk spacetimes induce boundary geometries which correspond to shearing motions of such a fluid. These motions have velocity profiles corresponding to regions of the flow field where the classical Kelvin--Helmholtz instability cannot arise; and yet the bulk physics is unstable when a characteristically string-theoretic effect, involving branes, is taken into account. However, it is not clear that the plasma endures for a sufficiently long time for this latter effect to become important observationally.

Fluid instability normally represents a transition from laminar to turbulent flow. Let us suppose that the QGP \emph{does} make such a transition at high $\mu$, before it hadronises, in a manner describable dually by the Seiberg--Witten effect. Then, as we have seen, this should occur throughout the fluid, and not merely deep in the fluid field. This might well have important consequences: for example, it could affect the ``spectator'' nucleons, that is, the nucleons in a peripheral collision adjacent to but not in the interaction zone.

Clearly one would like to gain a better understanding of this instability. One could try to use the $\ell$QKMV$_0$ geometry to study how perturbations propagate in the field theory on the boundary; this might help to identify the precise mechanism underlying the loss of stability. More ambitiously, one could try to model a turbulent flow arising from shear in the QGP by studying the back-reaction on the black hole geometry generated by branes as the Seiberg--Witten pair-production gets out of control. This might be feasible using the techniques described in \cite{kn:turbulent1,kn:turbulent2,kn:chesyaf}.

\addtocounter{section}{1}
\section*{\large{\textsf{Acknowledgements}}}
BMc is very grateful for the kind hospitality of NORDITA, for providing facilities and an environment in which this work could be initiated, and wishes to thank particularly L$\acute{\m{a}}$rus Thorlacius and the organisers of the ``Holographic Way'' programme. He also wishes to thank Prof.~Soon Wanmei for encouragement.  ET wishes to acknowledge the kind hospitality of the Centre for Gravitational Physics at The Australian National University, where this work was initiated.

\addtocounter{section}{1}
\section*{\large{\textsf{APPENDIX A: The Griffiths--Podolsk\'y metric}}}

In \cite{kn:grifpod,kn:exact}, Griffiths and Podolsk\'y showed how the Pleba\'nski--Demia\'nski solution \cite{kn:plebdem} can be rewritten in a more useful (albeit less symmetrical) form suitable for identifying the various different limits of this solution. In this appendix, we shall briefly review their form of this solution.

In its most general form, the Griffiths--Podolsk\'y metric contains eleven parameters, although only seven of them actually have physical significance. In the context of topologically spherical black holes, these seven parameters have the interpretation of mass ($m$), electric ($e$) and magnetic charge ($g$), angular momentum ($a$), NUT charge ($\ell$), acceleration ($\alpha$), and cosmological constant of the spacetime ($\Lambda$). Of the remaining four parameters, two ($k$,$\varpi$) are arbitrary scaling parameters, while two ($\epsilon$,$n$) are determined in terms of the rest when certain conditions are imposed on the nature of the solution.

For simplicity, we shall set the magnetic charge and acceleration parameters to zero. The resulting Griffiths--Podolsk\'y metric is\footnote{Note that we have set  $m=4\pi M^*$, $e=\sqrt{4\pi}\,Q^*$, and $\frac{\Lambda}{3}=-\frac{1}{L^2}$, but otherwise we generally follow the notation of \cite{kn:grifpod,kn:exact}.}
\ba
\label{GP1a}
g({\rm GP})&=&-\frac{\cal Q}{\rho^2}\left[\dif
t-\big(a(1-p^2)+2\ell(1-p)\big)\,\dif\phi\right]^2+\frac{\rho^2}{\cal Q}\,\dif
r^2\nonumber\\
&&+\frac{\rho^2}{{\cal P}}\,\dif
p^2+\frac{{\cal P}}{\rho^2}\left[{a}\,\dif
t-\big(r^2+(\ell+a)^2\big)\,\dif\phi\right]^2,
\ea
where
\ba
\label{GP1b}
\rho^2&=&r^2+(\ell+ap)^2\cr
{\cal P}&=&a_0+a_1p+a_2p^2+a_3p^3+a_4p^4\cr
{\cal Q}&=&\varpi^2k+4\pi Q^*{}^2-8\pi M^*r+\epsilon r^2+\frac{r^4}{L^2},
\ea
and
\ba
\label{GP1c}
a_0&=&\frac{1}{a^2}\Big(\varpi^2k+2nl-\epsilon\ell^2+\frac{\ell^4}{L^2}\Big)\cr
a_1&=&\frac{2}{a}\Big(n-\epsilon\ell+\frac{2\ell^3}{L^2}\Big)\cr
a_2&=&-\epsilon+\frac{6\ell^2}{L^2}\cr
a_3&=&\frac{4a\ell}{L^2}\cr
a_4&=&\frac{a^2}{L^2}.
\ea
The corresponding electromagnetic potential is
\be
\label{GP1d}
{\cal A}=-\frac{Q^*r}{\rho^2}\left[\dif
t-\big(a(1-p^2)+2\ell(1-p)\big)\dif\phi\right].
\ee
In this case, we immediately see that $\varpi$ is a redundant parameter and we set it to unity via a rescaling of $k$.

As explained in \cite{kn:grifpod,kn:exact}, the character of this solution depends on the number of real roots of ${\cal P}$. The usual case of topologically spherical black holes corresponds to ${\cal P}$ having two real roots, whose values can be set to be $p=\pm1$. This will then determine $\epsilon$ and $n$ in terms of the other parameters. For the planar black holes considered in this paper, ${\cal P}$ actually does not have any roots, and a different condition was imposed in section 3 to obtain expressions for $\epsilon$ and $n$.

\addtocounter{section}{1}
\section*{\large{\textsf{APPENDIX B: Constraining $\boldsymbol\ell$ from the Bulk Point of View}}}
Throughout our discussion we have assumed that $\ell$ is bounded above: to be precise, we have assumed that $\ell^2/(|a|L)$ is strictly bounded above by unity. This is well-motivated on the boundary, but it seems overly restrictive in the bulk. Here we point out, however, that it can also be motivated by considering the corresponding gravitational instanton (see equations (\ref{EUCLID}), (\ref{EUCLIDMORE}) above).

If $\ell^2/(|a|L) > 1$, the relevant roots of $\Delta_{\psi}^{\rm E}$  are
\ba
\Psi_1&=&\frac{-\ell+\sqrt{\ell^2-aL}}{a}\cr
\Psi_2&=&\frac{-\ell+\sqrt{\ell^2+aL}}{a}.
\ea
Note that $\Psi_2$ is different from the quantity with that name in the main discussion, although $\Psi_1$ is the same. The rod structure of this gravitational instanton will again consist of three rods, located at $r=r_{\rm H}^{\rm E}$ and $\psi=\Psi_{1,2}$. The first two rods are the same as in the case considered in section 4.2, with directions $\varrho_1$, $\varrho_2$ given by (\ref{l1}), (\ref{l2}) respectively. The third rod at $\psi=\Psi_2$ now has the direction
\be
\varrho_3=\frac{L}{2\sqrt{\ell^2+aL}}\left(L{\partial_t}+{\partial_\zeta}\right).
\ee

As before, the pair $\{\varrho_1,\varrho_2\}$ generates a $U(1)\times U(1)$ isometry of the space in the $(t,\zeta)$ direction. But now we need to ensure that the pair $\{\varrho_2,\varrho_3\}$ also generates the {\it same\/} $U(1)\times U(1)$ isometry. This requires $\{\varrho_1,\varrho_2\}$ and $\{\varrho_2,\varrho_3\}$ to be related by a $GL(2,\mathbb{Z})$ transformation \cite{kn:Chen}, i.e.,
\be
\Big(\begin{array}{c}
\varrho_1\\
\varrho_2
\end{array}\Big)=
\Big(\begin{array}{cc}
\alpha&\beta\\
\gamma&\delta
\end{array}\Big)
\Big(\begin{array}{c}
\varrho_2\\
\varrho_3
\end{array}\Big),
\ee
where $\alpha$, $\beta$, $\gamma$, $\delta$ are integers satisfying $\alpha\delta-\beta\gamma=\pm1$. Substituting in the expressions for $\varrho_1$, $\varrho_2$, $\varrho_3$, we obtain
\ba
\alpha&=&-\frac{L^2}{2\sqrt{\ell^2-aL}}\,\frac{{\Delta^{\rm E}_{r}}'(r_{\rm H}^{\rm E})}{(r_{\rm H}^{\rm E})^2-\ell^2-aL}\\
\label{cond2}
\beta&=&\sqrt{\frac{\ell^2+aL}{\ell^2-aL}}\,\frac{(r_{\rm H}^{\rm E})^2-\ell^2+aL}{(r_{\rm H}^{\rm E})^2-\ell^2-aL}=\pm1
\ea
and $\gamma=1$, $\delta=0$. A trivial solution to the second condition (\ref{cond2}) is $a=0$. On the other hand, the only non-trivial solution it has is
\be
a=\pm\frac{\sqrt{\ell^4-(r_{\rm H}^{\rm E})^4}}{L}.
\ee
But this is not real, since we required in section 4.2 that $(r_{\rm H}^{\rm E})^2>\ell^2+|aL|$ in order for the metric to have the correct signature. So no well-defined gravitational instanton exists in this case. Thus, our constraint on $\ell$ is well-motivated from the bulk point of view as well as by the boundary physics.

\end{document}